\documentclass{iopjournal}
\usepackage{amsfonts}
\usepackage{amsmath}
\usepackage{booktabs}
\usepackage{multirow}

\begin{document}

\articletype{paper} 

\title{Autoregressive long-horizon prediction of plasma edge dynamics}

\author{Hunor Csala$^1$\orcid{0000-0003-2993-8759}, Sebastian De Pascuale$^2$\orcid{0000-0001-7142-0246}, Paul Laiu$^3$\orcid{0000-0002-2215-6968}, Jeremy Lore$^2$\orcid{0000-0002-9192-465X}, Jae-Sun Park$^2$\orcid{0000-0003-0871-7527} and Pei Zhang$^{1,*}$\orcid{0000-0002-8351-0529}}

\affil{$^1$Computational Sciences and Engineering Division, Oak Ridge National Laboratory, Oak Ridge, TN, USA}

\affil{$^2$Fusion Energy Division, Oak Ridge National Laboratory, Oak Ridge, TN, USA}

\affil{$^3$Computer Science and Mathematics Division, Oak Ridge National Laboratory, Oak Ridge, TN, USA}

\affil{$^*$Author to whom any correspondence should be addressed.}

\email{zhangp1@ornl.gov}


\begin{abstract}
Accurate modeling of scrape-off layer (SOL) and divertor-edge dynamics is vital for designing plasma-facing components in fusion devices.
High-fidelity edge fluid/neutral codes such as SOLPS-ITER capture SOL physics with high accuracy, but their computational cost limits broad parameter scans and long transient studies. We present transformer-based, autoregressive surrogates for efficient prediction of 2D, time-dependent plasma edge state fields. Trained on SOLPS-ITER spatiotemporal data, the surrogates forecast electron temperature, electron density, and radiated power over extended horizons. 
We evaluate model variants trained with increasing autoregressive horizons (1--100 steps) on short- and long-horizon prediction tasks. Longer-horizon training systematically improves rollout stability and mitigates error accumulation, enabling stable predictions over hundreds to thousands of steps and reproducing key dynamical features such as the motion of high-radiation regions. Measured end-to-end wall-clock times show the surrogate is orders of magnitude faster than SOLPS-ITER, enabling rapid parameter exploration.
Prediction accuracy degrades when the surrogate enters physical regimes not represented in the training dataset, motivating future work on data enrichment and physics-informed constraints.
Overall, this approach provides a fast, accurate surrogate for computationally intensive plasma edge simulations, supporting rapid scenario exploration, control-oriented studies, and progress toward real-time applications in fusion devices.

\end{abstract}

\section{Introduction}
Tokamak magnetic confinement fusion devices face the challenge of requiring extremely high temperatures in the core while simultaneously 
maintaining heat and particle fluxes to specially armored divertor regions below the engineering limits. Divertor components are designed to effectively manage the intense heat flux and particle exhaust for plasma durations that enable reliable, safe fusion power plant operation. 
The open field-line scrape-off layer (SOL) region is the interface between the core and the plasma facing components. Accurate understanding and modeling SOL and divertor physics therefore become essential in reactor design and real-time control~\cite{kukushkin2011finalizing,wiesen2017plasma,senichenkov2022solps, ravensbergen2021real}. 

A range of physics-based tools exists for SOL and divertor modeling, spanning analytical two-point models~\cite{stangeby2000plasma}, reduced one-dimensional (1D) models (e.g., DIV1D~\cite{derks2022benchmark}), and high-fidelity two-dimensional (2D) simulations (e.g.,  SOLPS-ITER coupled with EIRENE Monte Carlo neutral model~\cite{wiesen2015new, bonnin2016presentation, park2024full} ). 
The SOLPS-ITER code provides a detailed description of plasma transport and plasma–neutral interactions in the divertor region, making it a key tool for reactor design studies. However, such simulations are computationally intensive, often requiring days to weeks of runtime for a single case. This high cost limits rapid exploration of the wide engineering design space, operational scenario development, and the evaluation of plasma exhaust mitigation strategies.
Consequently, reduced models are fast but often omit key physics, while high-fidelity simulations are typically too slow for real-time control and large parameter scans.

Deep learning (DL) has emerged as an attractive tool for plasma science~\cite{anirudh20232022} for its capability to capture complex, high-dimensional features from data. When trained on solutions from high-fidelity simulations, those DL-based approaches can deliver fast solutions while retaining much of the underlying accuracy. Wiesen et al.~\cite{wiesen2024data} summarize recent progress in integrating AI to advance understanding of plasma exhaust dynamics and plasma-edge processes.
In parallel, neural-operator approaches have been applied to solve magnetohydrodynamic (MHD) systems~\cite{gopakumar2024plasma,carey2025neural}, where Fourier Neural Operators (FNOs)~\cite{li2020fourier} are trained in an autoregressive manner to learn 2D temporal evolution. More recently, Paischer et al.~\cite{paischer2025gyroswin} demonstrated the potential of transformer architectures in high-dimensional plasma turbulence modeling with a Swin-based 5D gyrokinetic surrogate.

In the context of SOL and divertor physics, substantial progress has been made in developing DL surrogate models.
Existing work can be broadly grouped into three categories, trading off interpretability, speedup, and expressiveness. 
First, deep neural network (DNN) surrogates are used to replace the computationally intensive components within physics-based codes. This preserves the underlying physics and interpretability, but typically yields only moderate overall speedup. For example, Zhang et al.~\cite{zhang2025calculation} accelerate edge plasma simulations by introducing a transformer-based neural network (NN) surrogate for neural transport---the primary computational bottleneck in B2.5-EIRENE. Replacing the EIRENE Monte Carlo solver with this surrogate yields an 80-90\% reduction in runtime for B2.5-NN hybrid simulations. 
Second, reduced-order models can be constructed using user-specified basis functions to represent plasma dynamics, for instance via sparse identification of nonlinear dynamics (SINDy)~\cite{brunton2016discovering}. 
Lore et al.~\cite{lore2023time} use SINDy to develop reduced models for time-dependent SOLPS-ITER solutions suitable for model predictive control. Despite their interpretability, transparency, and strong performance, these approaches can be limited in expressiveness and struggle with complex, high-dimensional problems. For example, the model in~\cite{lore2023time} focuses on only the dynamics of two key parameters along the separatrix: the electron density at the outboard midplane and the electron temperature at the outer divertor. 
Third, for greater expressiveness and speedup, recent work trains DNN surrogates to directly predict SOL and divertor plasma states. Dasbach and Wiesen~\cite{dasbach2023towards} train feed-forward NNs and gradient-boost regression trees to map machine, operation, and transport parameters to the 1D outer-target electron temperature profile. Similarly, Li et al.~\cite{li2025multi} compare a fully connected NN and a convolutional NN (CNN) for predicting divertor target-plate particle flux density, target-plate electron temperature, and core-edge effective ion charge, reporting improved performance with CNN. 
Gopakumar and Samaddar~\cite{gopakumar2020image} train a fully convolutional network to map 2D edge plasma and neutral fields around a perturbed SOLPS-ITER steady state to future states 2 ms later. 
Zhu et al.~\cite{zhu2025latent} develop DivControlNN, a fast surrogate for steady-state divertor plasma predictions aimed at real-time detachment control. DivControlNN uses a two-stage approach: a multi-modal \(\beta\)-variational autoencoder (\(\beta\)-VAE) learns a latent representation of plasma diagnostics, and a subsequent multilayer perceptron (MLP) maps control parameters to this latent space, which is then decoded to predict diagnostics. Trained on a KSTAR database consisting of UEDGE solutions across a diverse parameter range, DivControlNN achieves quasi-real-time predictions with \(<20\)\% relative error and shows early success in detachment control. 

Much of the prior work, however, has focused on simplified settings---such as two-point dynamics \cite{lore2023time}, 1D representations~\cite{dasbach2023towards,poels2023fast,li2025multi,holt2024tokamak}, or steady-state~\cite{zhu2022data,zhu2025latent} and near steady-state conditions~\cite{gopakumar2020image}. 
These simplifications make it feasible to generate large, high-quality training datasets using established simulation codes, such as SOLPS-ITER, DIV1D, UEDGE, or Hermes-3. However, models trained under these assumptions may miss key physics and can yield inaccurate predictions and suboptimal design or control, as discussed in \cite{zhu2025latent}.
This gap motivates our work to develop a surrogate that predicts fully 2D, time-dependent plasma states. The work most closely related to ours is Poels et al.~\cite{poels2023fast}, which uses an autoregressive approach to predict transient dynamics, but focuses on 1D heat-flux tube between the X-point and the divertor target.

We propose a vision transformer (ViT)-based surrogate for time-dependent SOL plasma dynamics. Using global attention, the model captures long-range spatial correlations and multiscale features that are crucial to scrape-off-layer transport yet challenging to represent with approaches such as CNNs and SINDy. Related transformer-based surrogates have also been applied to other physical systems \cite{zhang2024matey, yin2025pixel}. We demonstrate that our surrogate delivers accurate and efficient predictions of 2D SOLPS-ITER fields while remaining stable over long autoregressive rollouts. These results open a pathway for integrating transformer-based architectures into fusion edge modeling.

The remainder of this paper is organized as follows. Section~\ref{sec:methods} introduces the proposed method, detailing SOLPS-ITER data generation, the DL problem formulation, the model architecture, and the training and testing setup. Section~\ref{sec:results} presents the prediction results and analyzes spatiotemporal behavior and error statistics. Section~\ref{sec:conc-dis} summaries the main findings, discusses the strengths and limitations of the proposed approach, and outlines directions for future work.

\section{Methods}
\label{sec:methods}

\subsection{Data generation}
\subsubsection{SOLPS-ITER setup} 

KSTAR plasmas are simulated using the SOLPS-ITER  code~\cite{wiesen2015new,bonnin2016presentation}. The SOLPS-ITER transport model represents the plasma using a fluid approximation with collisional parallel transport and an ad-hoc diffusive-convective cross-field representation. The plasma-neutral and plasma-surface interactions are handled using the kinetic Monte-Carlo code EIRENE. 

\begin{figure}
    \centering
    \includegraphics[width=0.9\linewidth]{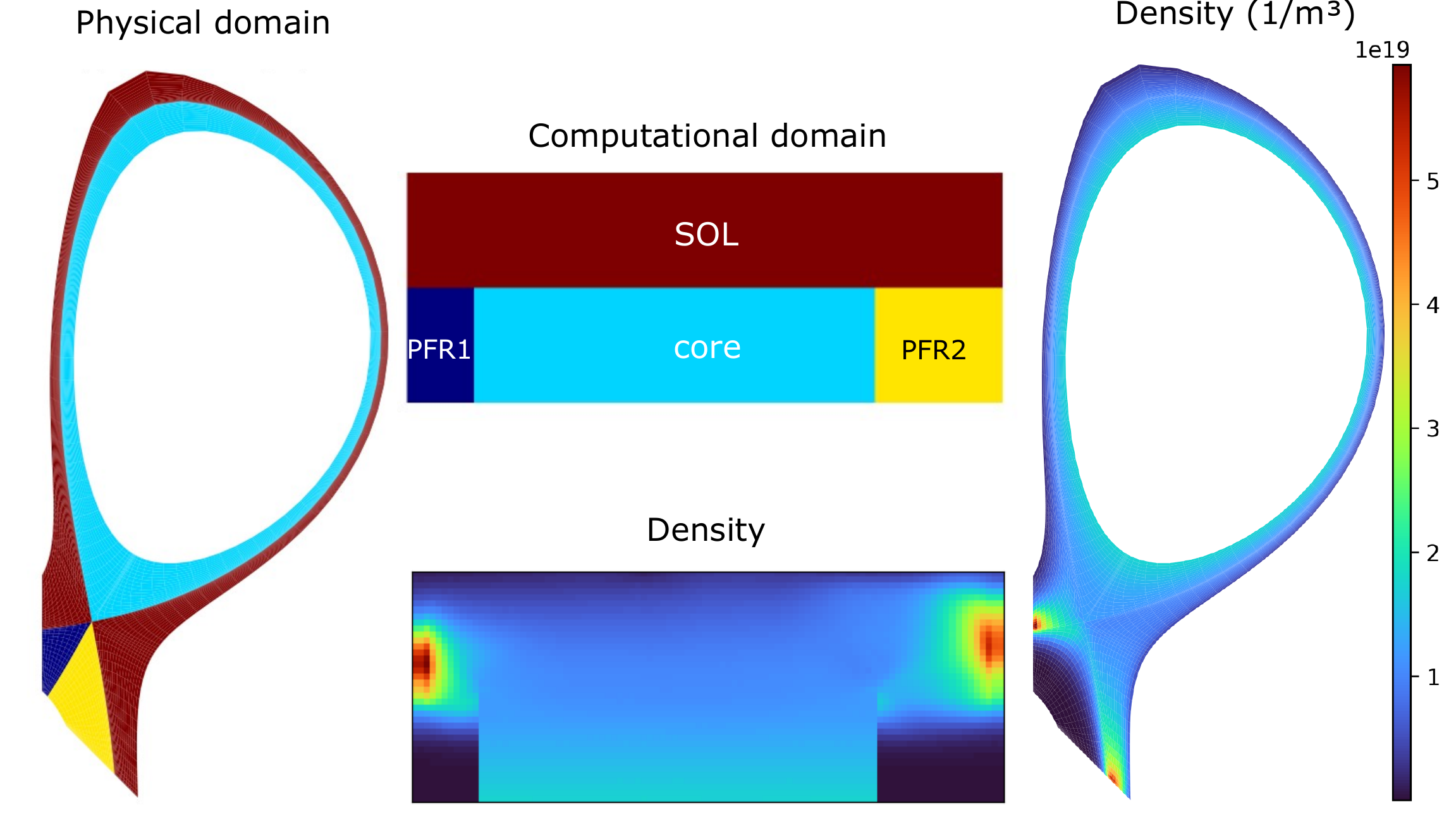}
    \caption{Schematic of the poloidal physical domain and its mapping to a rectangular computational domain, with regions color-coded. The scrape-off layer (SOL) is shown in red, the core in cyan, and the two private flux regions (PFRs) in the blue and yellow, respectively. For inference, an example density field is shown in both coordinate systems.}
    \label{fig:domain}
\end{figure}

KSTAR is a superconducting tokamak fusion device operated in Korea for magnetic confinement fusion research. The magnetic equilibrium is taken from the KSTAR L-mode detachment experiment \#19077, analyzed in Ref. \cite{park2018atomic}; the detailed physics settings and boundary conditions for this simulation here follow those documented in that work. The 2D structured computational grid is aligned with magnetic flux surfaces and consists of a rectangular mesh with 98$\times$38 cells, corresponding to the poloidal ($w$) and radial ($h$) directions, respectively. One layer of guard cells is included at each boundary of the computational domain. This numerical grid is mapped onto the physical plasma geometry, and the domain is divided into several regions: core in cyan, SOL in red, and two private flux regions (PFRs) in blue and yellow, as shown in Figure~\ref{fig:domain}. An example density field is shown in both coordinate domains for reference. 

The finite-volume numerical implementation is performed on a rectangular mesh using metric coefficients which capture the mapping to the physical-space coordinates of the KSTAR magnetic and plasma facing component topology. This mapping is convenient for data-driven applications, as the domain can be taken as a regular 2D image, instead of a complex graph structure. Our ML model operates entirely on the rectangular grid and does not incorporate information about the physical geometry; the mapping back to the physical domain is used solely for visualization.

\subsubsection{Three trajectories}\label{sec-threetraj}
We consider three transient 2D SOLPS-ITER simulations, each corresponding to a different actuator gas-puff rate history (Figure~\ref{fig:trajectories}). Trajectory~1 features a linearly ramped actuator input that spans a wide range of gas puff rate magnitudes, whereas trajectories~2 and 3 exhibit a more complex behavior, combining linear and sinusoidal components.
Depending on the waveform of the actuator trajectory, the resulting plasma response can differ substantially. Typical parallel transport timescales in KSTAR are on the order of a few tens of milliseconds \cite{park2023bifurcation}. When the actuator is varied at timescales comparable to, or faster than, this plasma response time, or when its amplitude changes more abruptly, the system is driven further away from quasi-steady-state (QSS) behavior. As a result, the three actuator trajectories here are expected to induce plasma states with different degrees of deviation from QSS conditions.
The SOLPS-ITER simulations carried out here are advanced with a time step of $\delta t = 0.1$~ms to generate sufficiently long trajectories with modest numerical noise, which in recent work has been shown to be correlated with time step and number of Monte Carlo particles \cite{vanuytven2025time}. We save the plasma state as snapshots every 10 time steps, corresponding to $\Delta t =1$~ms of physical time. Trajectories~1, 2, and 3 contain $N=908$, $4015$, and $7008$ snapshots, respectively.
Unless otherwise stated, we use the term \textit{step} to denote one snapshot interval, i.e., $\Delta t=1$~ms, for the rest of the paper. 
For snapshot time steps labeled as \(t=0,\ldots, N-1\), the state is stored as a tensor \(\mathbf{u}_t \in \mathbb{R}^{H \times W \times C}\), where \(H = 38\) and \(W = 98\) are respectively the radial and poloidal grid resolutions and \(C\) is the number of physical variables (channels) included in the plasma state. 
In this work, the variables include the electron density, electron temperature, and total radiated power ($C=3$).

\subsection{Problem setting}

\subsubsection{Autoregressive prediction task}
Our goal is to develop a surrogate model that can autoregressively capture 2D plasma dynamics driven by a time-varying actuator input (the gas puffing rate signal). Specifically, given the previous $T$ plasma states and the actuator history up to the current time, together with the next actuator input, the model predicts the next plasma state (one step ahead). This prediction is then fed into the model to roll the system forward in time in an autoregressive manner given an actuator input series. 
Specifically, at time step $t$, the autoregressive approach predicts the next $n_{\mathrm{lead}}$ plasma states from last $T$ plasma states $[\mathbf{u}_{t-T+1},\dots,\mathbf{u}_{t-1},\mathbf{u}_{t}]$ and actuator inputs $[a_{t-T+1},\dots,a_t,\dots, a_{t+n_{\mathrm{lead}}}]$ by performing
\begin{equation}
    \label{eq:model_auto}
    \begin{aligned}
    \mathbf{\hat{u}}_{t+1}&=f_{\theta}([\mathbf{u}_{t-T+1},\dots,\mathbf{u}_{t-1},\mathbf{u}_{t}];[a_{t-T+1},\dots,a_{t}, a_{t+1}]),\\
    \mathbf{\hat{u}}_{t+2}&=f_{\theta}([\mathbf{u}_{t-T+2},\dots,\mathbf{u}_{t},\mathbf{\hat{u}}_{t+1}];[a_{t-T+2},\dots,a_{t+1}, a_{t+2}]), \\
    &\vdots \\
    \mathbf{\hat{u}}_{t+n_{\mathrm{lead}}}&=f_{\theta}([\mathbf{\hat{u}}_{t-T+n_{\mathrm{lead}}},\dots,\mathbf{\hat{u}}_{t+n_{\mathrm{lead}}-2},\mathbf{\hat{u}}_{t+n_{\mathrm{lead}}-1}];[a_{t-T+n_{\mathrm{lead}}},\dots,a_{t+n_{\mathrm{lead}}-1},a_{t+n_{\mathrm{lead}}}]),
\end{aligned}
\end{equation}
where \(\mathbf{\hat{u}}_s\) is the predicted plasma state at time step $s=t+1,\dots,t+n_{\mathrm{lead}}$ given by the surrogate model \(f_{\theta}\) with parameters \(\theta\) learned during the training procedure. 
Tasks of this type are common for modeling spatiotemporal evolution of partial differential equations (PDEs)~\cite{gopakumar2024plasma, carey2025neural, herde2024poseidon, mccabe2025walrus}. 

\subsubsection{Database}
\begin{figure}
    \centering
    \includegraphics[width=0.7\linewidth]{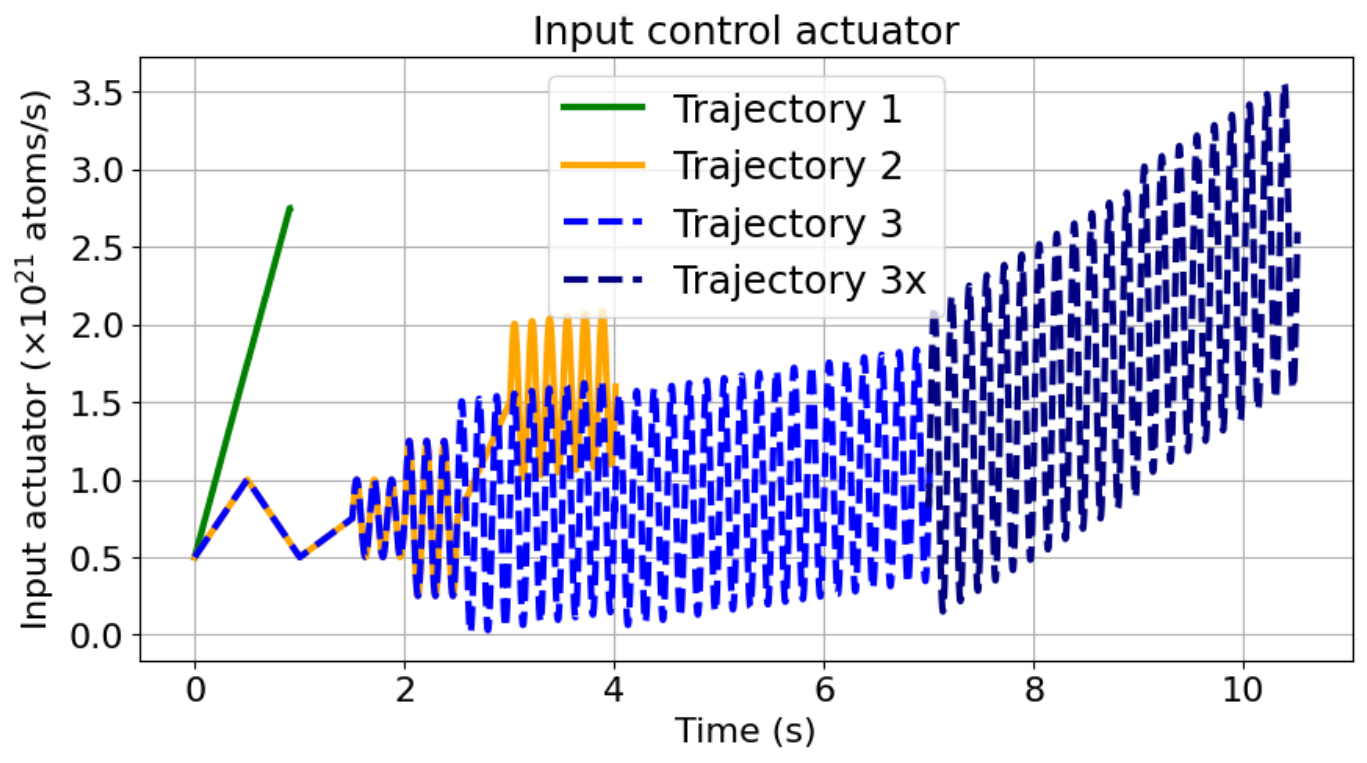}
    \caption{Input gas-puff trajectories (actuator signals) used in this study, shown as a function of time. Trajectories~1 and 2 are used for training, while trajectory~3 is reserved for testing.
    Trajectory~3x extends trajectory~3 beyond 7~s (with increased puffing rates) and is used as a separate, more challenging stress-test case.
}
    \label{fig:trajectories}
\end{figure}

Our database comprises spatiotemporal 2D solutions from trajectories 1--3 (Figure~\ref{fig:trajectories}; Section~\ref{sec-threetraj}). Trajectory~1 applies a purely linear actuator gas-puff signal with the largest amplitudes. Trajectories~2 and 3 superimpose linear and sinusoidal components and remain overlapping until approximately 2.5~s. 
In this study, we train the surrogates using trajectories~1 and 2, and assess generalization on trajectory~3.  
We also consider a more challenging case for evaluation---trajectory 3x, which extends trajectory~3 from 7~s to 10.5~s and puffing rates beyond 3.5~$\cdot 10^{21}$ atoms/second, with an extra 3507 time steps. With the extension, trajectory 3x results in a deep nonlinear plasma response that causes the plasma radiation front to move from the divertor area, along the separatrix leg, and to the x-point region. We will use it as a separate stress test of the model's performance.

\subsection{MATEY model}
We build the plasma surrogate model \(f_{\theta}\) using the MATEY codebase~\cite{zhang2024matey,yin2025pixel}. MATEY is a scalable, transformer-based PyTorch framework that has been used used to model a range of fluid-dynamics systems \cite{zhang2024matey,yin2025pixel}. It supports multiple spatiotemporal transformer variants spanning different degrees of factorization, from the fully decoupled AViT~\cite{mccabe2024multiple} to SViT, and the fully coupled ViT, as well as the more recent hierarchical multiscale Turbulence Transformer designed for extreme-resolution data~\cite{yin2025pixel}.

\begin{figure}
    \centering
    \includegraphics[width=0.9\linewidth]{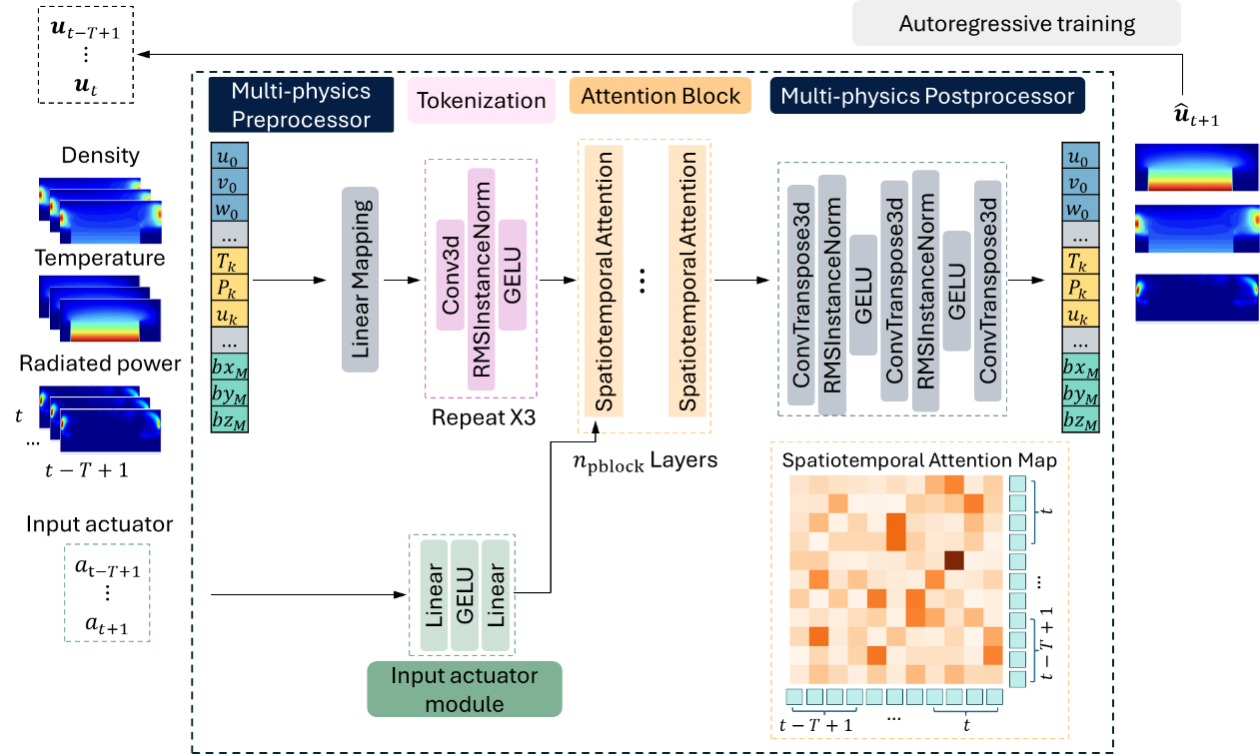}
    \caption{Schematic of the MATEY model architecture with ViT. The model takes as input the plasma states from the previous \(T\) time steps, \(\mathbf{U}_{t,T} =[\mathbf{u}_{t-T+1}, \dots, \mathbf{u}_{t}]\), together with the actuator values from the previous \(T\) time steps and the next timestep, \(\mathbf{a}_{t+1,T+1}=[a_{t-T+1}, \dots, a_{t}, a_{t+1}]\). It outputs the predicted plasma state at the next timestep, \(\hat{\mathbf{u}}_{t+1}\). Spatiotemporal attention enables each patch to attend to all other patches both across space and time steps (see bottom-right panel). The model is trained autoregressively, feeding predictions back as inputs for subsequent time steps.} 
    \label{fig:matey_model}
\end{figure}

In this work, we adopt the ViT backbone in MATEY, for the data resolution $H\times W=38\times98$. Full spatiotemporal attention is computationally tractable at these dimensions, which allows for capturing correlations across all spatiotemporal points. 
An overview of the model architecture is shown in Figure~\ref{fig:matey_model}. Given past plasma states \(\mathbf{U}_{t,T} =[\mathbf{u}_{t-T+1}, \dots, \mathbf{u}_{t}]\in \mathbb{R}^{T \times H \times W \times C}\), past actuator signals \(\mathbf{a}_{t,T}=[a_{t-T+1}, \dots, a_{t}]\in \mathbb{R}^T\), and the next control actuator \(a_{t+1}\), MATEY predicts the next-step plasma state \(\hat{\mathbf{u}}_{t+1}\) with the following major modules: 
\begin{itemize}
    \item \textit{Multi-physics preprocessor}. A linear projection maps the input multi-physics input tensor \(\mathbf{U}_{t,T}\in \mathbb{R}^{T \times H \times W \times C}\) into a unified representation \(\mathbf{U}_\text{uni}\in \mathbb{R}^{T \times H \times W \times C_\text{uni}}\). 
    
    \item \textit{Tokenization.} The unified field representation \(\mathbf{U}_\text{uni}\) is discretized into spatiotemporal tokens \(\mathbf{Z}^0\in \mathbb{R}^{L \times C_\text{emb}} \) for transformer processing. The tokenization module consists of a stack of convolutional blocks, resulting in the embedding dimension ${C_\text{emb}}$ and sequence length \(L=T/p_t\times H/p_h\times W/p_w\), where \((p_t,p_h,p_w)\) denotes the effective patch size.
    
    \item \textit{Input actuator module.} The puffing rate tensor \(\mathbf{a}_{t+1,T+1}\in \mathbb{R}^{T+1}\) is embedded with an MLP block to a high-dimensional representation \(\mathbf{h}_a\in \mathbb{R}^{C_\text{emb}}\).
    
    \item \textit{Attention module.} All-to-all correlation among the spatiotemporal tokens are modeled using \(L_{\mathrm{pblock}}\) spatiotemporal transformer blocks. Each of these blocks consists of a multi-head self-attention~\cite{vaswani2017attention,dosovitskiy2020image}, followed by an MLP. The input actuator representation \(\mathbf{h}_a\) is added to the first block. We employ global spatiotemporal attention over the full token sequence (see the spatiotemporal attention map in Figure~\ref{fig:matey_model}), so that each patch at each timestep can attend to all other patches across space and time. This is important because regions that are adjacent in the computational domain may be far apart in the physical domain (Figure~\ref{fig:domain}). Poels et al.~\cite{poels2023fast} address this by introducing a geometric mask to restrict attention to physically neighboring regions. In contrast, we impose no such constraint, allowing the model to learn relevant spatial relationships directly from the data. 
    
    \item \textit{Multi-physics postprocessor.} The final tokens representation from attention \(\mathbf{Z}^{L_{\mathrm{pblock}}}\in\mathbb{R}^{L\times C_\text{emb}}\) is decoded back to the physical field tensor \(\hat{\mathbf{u}}_{t+1}\in \mathbb{R}^{H \times W \times C}\) using a stack of transposed convolutional layers (Figure~\ref{fig:matey_model}).
\end{itemize}
In our experiments, we set \( T=3\), \(C_\text{uni}=48\), \(C_\text{emb}=192\), patch size \((p_t,p_h,p_w)=(1,2,2)\) and \(L_{\mathrm{pblock}} = 12\). To mitigate scale disparities among the input variables, each input channel and the actuator signal are independently normalized using min–max scaling to the range~\([0,1]\).

We train our model autoregressively, where the model predictions are fed back as inputs for subsequent steps (Eq.~\eqref{eq:model_auto}). The autoregressive training is crucial for robust long-horizon prediction. When the model is trained only for next-step prediction, the error accumulation over time steps results in inputs quickly drifting away from the training distribution during inference, which often leads to divergence in longer rollouts. By employing autoregressive training with \(n_{\mathrm{lead}}>1\), the error accumulation over time is suppressed in the training procedure, which mitigates the divergence issue. 
Backpropagating through consecutive model calls, however, can be prohibitively expensive for long rollouts. We adopt the pushforward trick of Brandstetter et al.~\cite{brandstetter2022message}, in which gradients are propagated only at the final step of the rollout.
Conceptually, this approach queries the model multiple times to generate a perturbed prediction, which is then used as the input in place of the ground truth state. This yields substantial computational savings while preserving rollout stability. A similar approach has been utilized in \cite{poels2023fast} for 1D divertor plasma predictions, and likewise helps maintain robustness during extended rollouts without incurring prohibitive cost.

For each training batch, we first sample starting times independently and uniformly from all time steps of trajectories 1 and 2. The rollout horizon is then uniformly sampled per batch from 1 to min(\(n_{\mathrm{lead}}, n_\mathrm{max}\)), where \(n_\mathrm{max}\) is the longest rollout allowed by the latest start time in the batch (i.e., the remaining number of steps in the trajectory). Sampling the rollout length uniformly encourages the model to learn both short- and long-horizon predictions, while inducing a mild preference for shorter rollouts when starting times are close to the trajectory end.

We train four models with \(n_{\mathrm{lead}}\in \{1, 10, 50, 100\}\) (with \(n_{\mathrm{lead}}=1\) for next-step prediction) to study the effect of rollout horizon on predictive performance: \textit{Matey-1}, \textit{Matey-10}, \textit{Matey-50}, \textit{Matey-100}. 
We use the normalized mean square error (NMSE) as the training loss and optimize by using the Adam-based DAdaptAdam optimizer~\cite{defazio2023learning} paired with a cosine annealing scheduler to adapt the effective learning rate during training. 
For DAdaptAdam, we set the learning rate scale to 1 and the growth rate to \(1.05\). 
The NMSE loss is computed by normalizing the MSE by the mean squared ground-truth values. Specifically, given a current time step $t$ and a prediction leadtime $n_{\mathrm{lead}}$, let \(t^\prime=t+n_{\mathrm{lead}}\). For ground truth \(\mathbf{u}_{t^\prime} \in \mathbb{R}^{H \times W \times C}\) and prediction \(\hat{\mathbf{u}}_{t^\prime} \in \mathbb{R}^{H \times W \times C}\), the loss is defined as
\begin{equation}\label{eq-mseloss}
   \text{Loss}(t, n_{\mathrm{lead}})=\frac{1}{C}\sum_c\frac{\sum_{h,w}{(\hat{\mathbf{u}}_{t^\prime} - \mathbf{u}_{t^\prime})^2}_{h,w,c}}{\sum_{h,w}(\mathbf{u}_{t^\prime})^2_{h,w,c}}.
\end{equation}
The loss is computed only at the final time \(t^\prime\). In the results section, we report the variable-averaged normalized root-mean-squared errors (NRMSE) as
\begin{equation}\label{eq-nrmse}
   \varepsilon(t, n_{\mathrm{lead}})=\frac{1}{C}\sum_c\sqrt{\frac{\sum_{h,w}{(\hat{\mathbf{u}}_{t^\prime} - \mathbf{u}_{t^\prime})^2}_{h,w,c}}{\sum_{h,w}{(\mathbf{u}_{t^\prime})}^2_{h,w,c}}},
\end{equation}
for a current (starting) timestep \(t\) and rollout (leadtime) steps \(n_{\mathrm{lead}}\) during inference.

All models are trained for 100,000 optimization steps with a batch size of 64, corresponding to approximately 1,300 epochs (full passes through the dataset). 
To stabilize early training, we cap the maximum rollout length at \(n_{\mathrm{lead}}/2\) for the first 1000 steps, which helps prevent divergence before the model has learned reliable short-term dynamics. 
The \textit{Matey-100} model is trained an additional 20,000 steps to reach convergence. Training is primarily performed on the Pittsburgh Supercomputing Center (PSC) Bridges-2 system~\cite{brown2021bridges} using NVIDIA V100 and H100 GPUs. Training was performed on 8 GPUs using distributed data parallelism. 
The final training losses are summarized in Table~\ref{tab:final_training_losses}.
Values report the average training loss over one or multiple runs, with standard deviations (STDs) included where available. 
For models \textit{Matey-1} and \textit{Matey-10}, the reported standard deviations are calculated from two and three runs, respectively, to quantify the effects of randomness in training and initialization.
The small standard deviations (two orders of magnitude smaller than the corresponding mean) indicate that training is robust across runs. As expected, the losses increase slightly for larger \(n_{\mathrm{lead}}\), reflecting the increased difficulty of learning longer autoregressive rollouts.

\begin{table}[t]
\centering
\begin{tabular}{l c}
\hline
Model & Final training loss \\
\hline
\textit{Matey-1}   & $0.004807 \pm 4.239\times 10^{-5}$ \\
\textit{Matey-10}  & $0.004971 \pm 1.596\times 10^{-5}$ \\
\textit{Matey-50}   & $0.005281$ \\
\textit{Matey-100} & $0.005747$ \\
\hline
\end{tabular}
\caption{Final training losses (mean $\pm$ standard deviation (STD) where available).}
\label{tab:final_training_losses}
\end{table}

\section{Results}
\label{sec:results}
In this section, we present results in four subsections:
\begin{itemize}
    \item Section~\ref{subsec:global_performance}: \textit{Global performance across trajectories 1--3.} We evaluate model accuracy on trajectories~1, 2, and 3 using the global error metric $\varepsilon(t,n_{\mathrm{lead}})$ (Eq.~\eqref{eq-nrmse}) for varying start times $t$ and leadtimes $n_{\mathrm{lead}}$. This analysis quantifies (i) the impact of training rollout lengths on the short- and long-horizon prediction performance and (ii) the model performance on the unseen trajectory~3 when trained on trajectories 1 and 2.
    \item Section~\ref{subsec:detailed_analysis_traj3}: \textit{Detailed analysis on trajectory 3.}  We examine trajectory 3 in more detail via (i) 2D contour comparisons of the plasma states, i.e., density, temperature, and radiated power; (ii) time-history comparisons at a selected upstream outer SOL cell located near the outboard midplane; and (iii) 1D spatial profile comparisons at selected locations. In addition to accuracy, we visualize attention maps to assess whether the model learns meaningful spatiotemporal correlations from data to provide qualitative insights into its predictions.
    \item Section~\ref{subsec:generalization_to_3x}: \textit{Generalization to the more challenging trajectory 3x.} We test the model on trajectory 3x to assess whether it can capture the peak location of radiated power across distinct physical regimes.
    \item Section~\ref{subsec:runtime}: \textit{Runtime.} We report the training and inference times.
\end{itemize}

\subsection{Global performance across trajectories 1--3}
\label{subsec:global_performance}

\begin{figure}
    \centering
    \includegraphics[width=\linewidth]{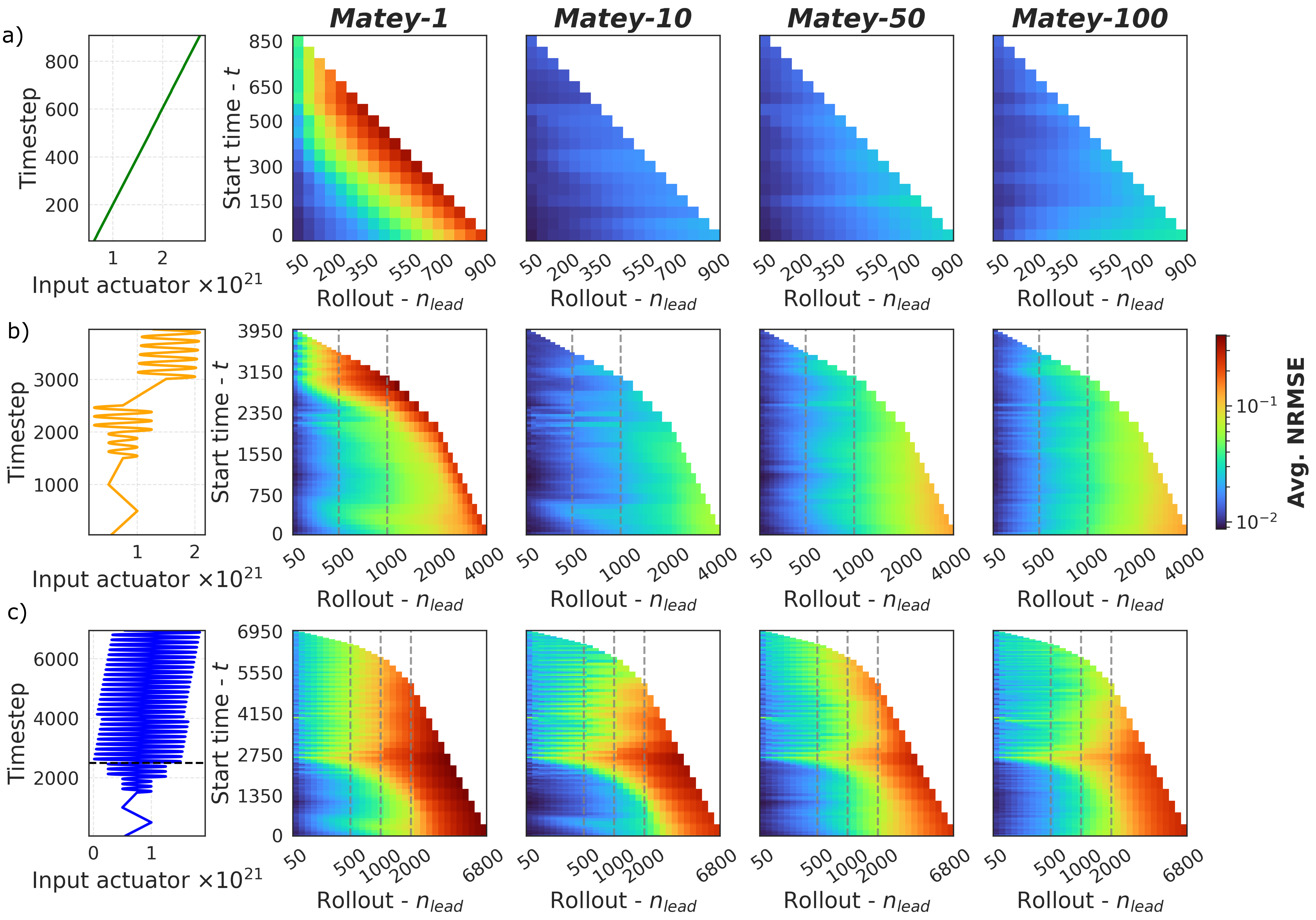}
    \caption{Prediction NRMSE \(\varepsilon(t, n_{\mathrm{lead}})\) (Eq. \eqref{eq-nrmse}) versus start time \(t\) and rollout horizon \(n_{\mathrm{lead}}\) for the four models, \textit{Matey-1}, \textit{Matey-10}, \textit{Matey-50}, and \textit{Matey-100}, for trajectories~1--3. Each row corresponds to a distinct trajectory, with the first column showing the associated input actuator signal. In panel c), the horizontal dashed black line indicates the point up to which trajectories 2 and 3 overlap. The subsequent columns display results from four models trained with different numbers of autoregressive steps \(n_{\mathrm{lead}}\). For panel b) and c), the horizontal axis of the heatmaps is non-uniform. Vertical gray lines denote changes in resolution: the first segment has a resolution of 50 time steps, and the resolution is halved after each gray line. Errors increase clearly with longer rollouts, with \textit{Matey-10} achieving the best accuracy on the training trajectories (panels a) and b)) and \textit{Matey-100} performing best on the test trajectory (panel c)). \textit{Matey-1}, which is trained with simple next-step prediction, results in highest error among all models. The prediction error also varies depending on the start time, as discussed in Section~\ref{subsec:global_performance}.}
    \label{fig:train_test_rollout_error}
\end{figure}

Figure~\ref{fig:train_test_rollout_error} shows the contour plots of NRMSE, \(\varepsilon(t, n_{\mathrm{lead}})\) (see Eq.~\eqref{eq-nrmse}), versus start time \(t\) and rollout horizon \(n_{\mathrm{lead}}\) for the four models, \textit{Matey-1}, \textit{Matey-10}, \textit{Matey-50}, and \textit{Matey-100}, on trajectories~1--3.
Each row corresponds to one trajectory with the corresponding input actuator signal shown in the first column. The black horizontal dashed line in panel c) marks the time up to which trajectories 2 and 3 overlap. The next four columns are the NRMSE contour plots for the four models, respectively, with a colormap on a logarithmic scale.

To generate the contours for trajectory 1, we use a uniform \((t, n_{\mathrm{lead}})\) grid with 18 points in each direction. For trajectories 2 and 3 (panels b) and c)) we use non-uniform grids with progressively coarse cells at larger \((t, n_{\mathrm{lead}})\) values to accommodate the longer trajectories. Vertical gray lines mark resolution changes: the first segment has a resolution of 50 time steps, and the resolution is halved after each gray line. Consequently, grid cells in the heatmaps become progressively larger, reaching 200 and 400 time steps for trajectories 2 and 3, respectively. After each resolution change, the vertical resolution is also halved with a starting resolution of 50 time steps. The heatmaps are only populated below the diagonal, since the sum of the start time and the number of rollout steps cannot exceed the trajectory length (\(t+n_{\mathrm{lead}}\leq N\)). Panels b) and c) have a non-uniform horizontal axis with a uniform vertical axis, which causes the diagonal to appear distorted.

From the error contours in Figure~\ref{fig:train_test_rollout_error}, \textit{Matey-1} (second column; trained using next-step predictions only) exhibits the poorest performance across all trajectories. As expected, the errors increase with rollout horizon (horizontal axis) for all three trajectories due to the error accumulation. 
Notably, for the two training trajectories (i.e., trajectories~1 and 2), \textit{Matey-1} also shows a pronounced error increase at later start times. This is likely due to a training-data bias: fewer training samples have the larger gas puff rates that occur at later times (Figure~\ref{fig:trajectories}).  
In contrast, this start-time dependence is largely mitigated in \textit{Matey-10}, \textit{Matey-50}, and \textit{Matey-100}, whose errors are more uniformly distributed along the vertical axis. Moreover, these models reduce the NRMSE for full-trajectory rollout from 40\% (in \textit{Matey-1}) to below 10\% for the two training trajectories. This suggests that incorporating autoregressive training helps the models better capture the dynamics and reduce sensitivity to data imbalance. 

For trajectory~3 in the third row, a distinct jump in error appears around 2500 steps, corresponding to the end of the overlapping region between trajectories~2 and~3. Since trajectory~2 is included during training, this point marks the transition from the training regime to the testing regime, which is unseen during training. 
Across \textit{Matey-1}--\textit{Matey-100} (from left to right in panel c) of Figure~\ref{fig:train_test_rollout_error}), longer-rollout training improves long-horizon accuracy, especially in the test regime (\(t>2500\)), while all models remain accurate for short horizons (50–100 steps). Overall, \textit{Matey-10} achieves the best performance within the training regime, while \textit{Matey-50} and \textit{Matey-100} extrapolate better with lower errors in the testing regime. 

To provide a quantitative comparison, Table~\ref{tab:errors} reports the mean and standard deviation (mean\(\pm\)STD), computed over the start time \(t\)), of the NRMSE at selected rollout horizons \(n_\mathrm{lead}\) for the contours in Figure~\ref{fig:train_test_rollout_error}. For trajectory~3, the statistics are computed over \(t>2500\) to focus on the unseen dynamics during training.
\textit{Matey-1} degrades sharply as the rollout horizon increases and exhibits large STDs---most notably on trajectory~1---reflecting strong sensitivity to the start time. In contrast, the other three models maintain small STDs, indicating robust and consistent performance over different start times. 
Consistent with the contour plots, \textit{Matey-10} yields the lowest mean NRMSE on the training trajectories~1 and~2, whereas \textit{Matey-50} and \textit{Matey-100} achieve the lowest mean NRMSE on the test trajectory~3, suggesting improved generalization to previously unseen dynamics. For the previously unseen dynamics with longer horizons (trajectory~3, \(n_\mathrm{lead} \ge 500\)), \textit{Matey-100} clearly outperforms the others, with both lower mean and STDs, highlighting the benefits of long-horizon training.

\begin{table}[t]
\centering
\small
\resizebox{\textwidth}{!}{
\begin{tabular}{c l c c c c}
\toprule
\textbf{Trajectory} &  & \textbf{Matey-1} & \textbf{Matey-10} & \textbf{Matey-50} & \textbf{Matey-100} \\
& Rollout \(n_\mathrm{lead}\) &  &  &  & \\
\midrule
 & 100 & 3.82e-2$\pm$2.54e-2 & \textbf{1.14e-2$\pm$9.63e-4} & 1.18e-2$\pm$9.85e-4 & 1.24e-2$\pm$8.21e-4 \\
\textbf{1} & 500 & 1.46e-1$\pm$1.02e-1 & \textbf{1.58e-2$\pm$1.22e-3} & 1.93e-2$\pm$1.81e-3 & 2.06e-2$\pm$1.89e-3 \\
 & 700 & 1.78e-1$\pm$7.34e-2 & \textbf{1.81e-2$\pm$9.59e-4} & 2.38e-2$\pm$2.13e-3 & 2.53e-2$\pm$1.84e-3 \\ \midrule
 & 100 & 1.78e-2$\pm$9.22e-3 & \textbf{1.10e-2$\pm$1.31e-3} & 1.12e-2$\pm$9.46e-4 & 1.18e-2$\pm$1.11e-3 \\
 & 500 & 5.58e-2$\pm$4.86e-2 & \textbf{1.61e-2$\pm$2.40e-3} & 1.99e-2$\pm$2.03e-3 & 2.19e-2$\pm$2.38e-3 \\
\textbf{2} & 1000 & 9.32e-2$\pm$9.16e-2 & \textbf{2.24e-2$\pm$2.09e-3} & 3.30e-2$\pm$2.28e-3 & 3.64e-2$\pm$2.27e-3 \\
 & 2000 & 1.14e-1$\pm$5.56e-2 & \textbf{3.49e-2$\pm$2.26e-3} & 6.28e-2$\pm$2.14e-3 & 6.59e-2$\pm$2.13e-3 \\
 & 3200 & 1.68e-1$\pm$3.65e-2 & \textbf{5.19e-2$\pm$1.60e-3} & 1.03e-1$\pm$2.77e-3 & 1.02e-1$\pm$3.32e-3 \\
 \midrule
 & 100 & 2.55e-2$\pm$5.13e-3 & 2.43e-2$\pm$6.08e-3 & \textbf{2.17e-2$\pm$5.44e-3} & 2.37e-2$\pm$5.53e-3 \\
 & 500 & 6.61e-2$\pm$1.05e-2 & 4.75e-2$\pm$1.66e-2 & 4.35e-2$\pm$1.00e-2 & \textbf{3.83e-2$\pm$1.04e-2} \\
\textbf{3} & 1000 & 1.20e-1$\pm$1.30e-2 & 8.09e-2$\pm$2.55e-2 & 7.42e-2$\pm$1.57e-2 & \textbf{5.69e-2$\pm$1.40e-2} \\
 & 2000 & 2.17e-1$\pm$1.67e-2 & 1.54e-1$\pm$3.30e-2 & 1.32e-1$\pm$2.06e-2 & \textbf{9.99e-2$\pm$1.81e-2} \\
 & 4000 & 3.76e-1$\pm$1.17e-2 & 3.05e-1$\pm$2.41e-2 & 2.50e-1$\pm$9.35e-3 & \textbf{2.03e-1$\pm$6.87e-3} \\
\bottomrule
\end{tabular}}
\caption{Mean \(\pm\) STD of NRMSE, aggregated over start time \(t\), at selected rollout horizons \(n_{\mathrm{lead}}\) for each trajectory and model in Figure~\ref{fig:train_test_rollout_error}. For trajectory~3, aggregation is performed over  \(t>2500\) for the unseen dynamics only.
Larger STDs indicate greater sensitivity to the start time. \textit{Matey-1} exhibits largest variability across \(t\). \textit{Matey-10} and \textit{Matey-50} perform the best on the training trajectories~1 and 2, whereas \textit{Matey-100} achieves the lowest errors on the unseeen test trajectory~3 for longer rollout horizons. 
}
\label{tab:errors}
\end{table}

\subsection{Detailed analysis on trajectory 3}
\label{subsec:detailed_analysis_traj3}

\subsubsection{2D spatial distribution prediction}

\begin{figure}
    \centering
    \includegraphics[width=0.6\linewidth]{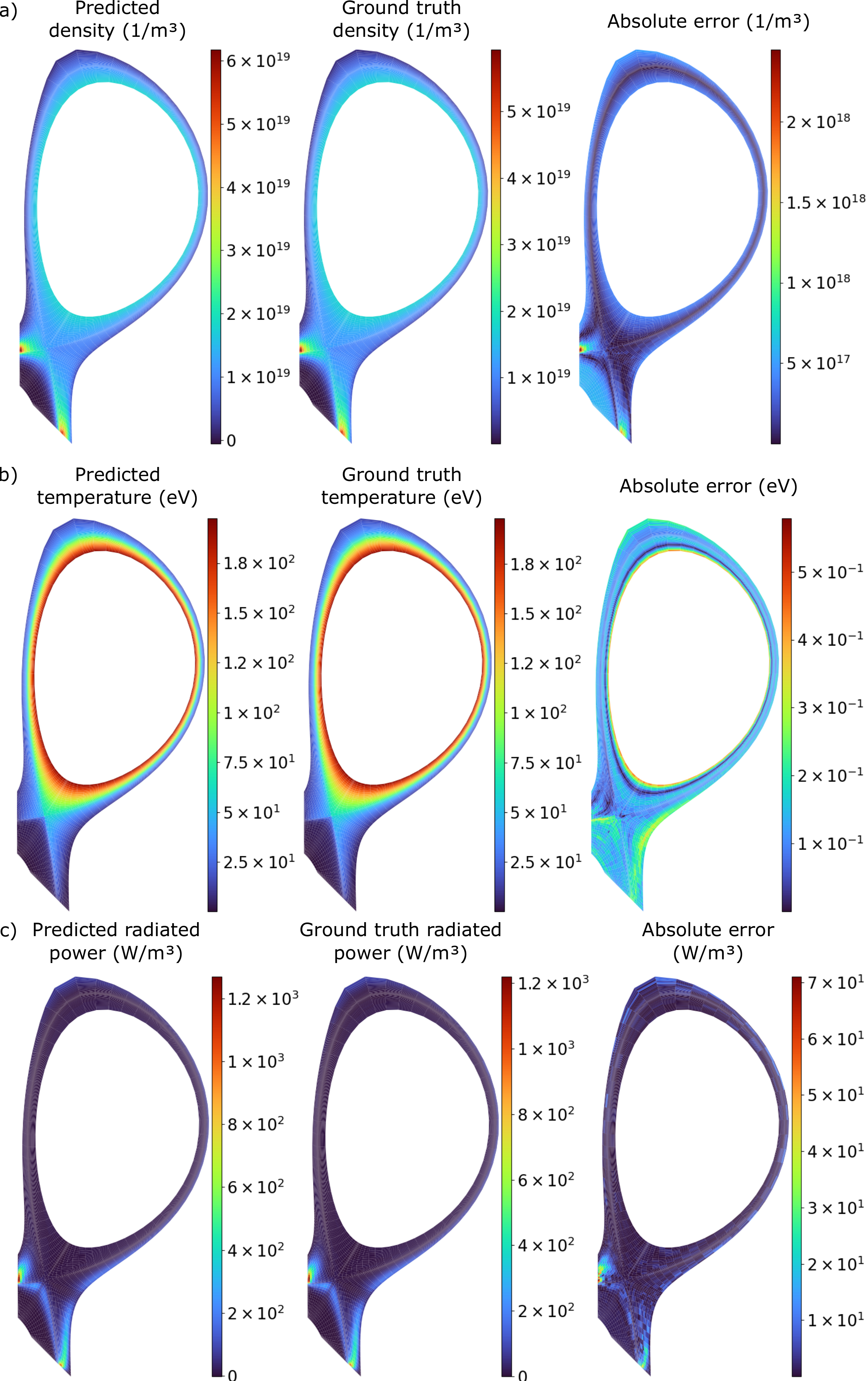}
    \caption{Comparison of density, temperature, and radiated power between the \textit{Matey-100} prediction and ground truth from SOLPS. Results are shown for Trajectory 3 at \(t=7\) s, after 100 rollout steps from \(6.9\) s. The last column shows the absolute errors. Predictions align closely with the ground truth, with errors one to two orders of magnitude smaller than the corresponding field values.}
    \label{fig:2D_errors}
\end{figure}

Figure~\ref{fig:2D_errors} compares \textit{Matey-100} predictions with the ground-truth fields of density, temperature, and radiated power mapped onto the physical domain. The final column displays the corresponding absolute error fields. Results are shown for trajectory~3 at \(t=7\) s, after 100 rollout steps starting from 6.9 s. Across all variables, the absolute errors are typically one to two orders of magnitude smaller than the respective ground-truth values, indicating strong predictive performance. Density exhibits the highest relative errors overall, with relatively large discrepancies in the divertor, coinciding with high-density regions. For temperature, errors peak along the inner edge of the domain, whereas low errors are observed near the inner side of the divertor region and moderate errors near the outer edge. For radiated power, errors are more pronounced in the inner and outer divertor regions. Overall, the model produces consistently accurate predictions of key plasma quantities across the full 2D domain.

\subsubsection{Time history at the outer SOL upstream near the outboard midplane}

\begin{figure}
    \centering
    \includegraphics[width=0.85\linewidth]{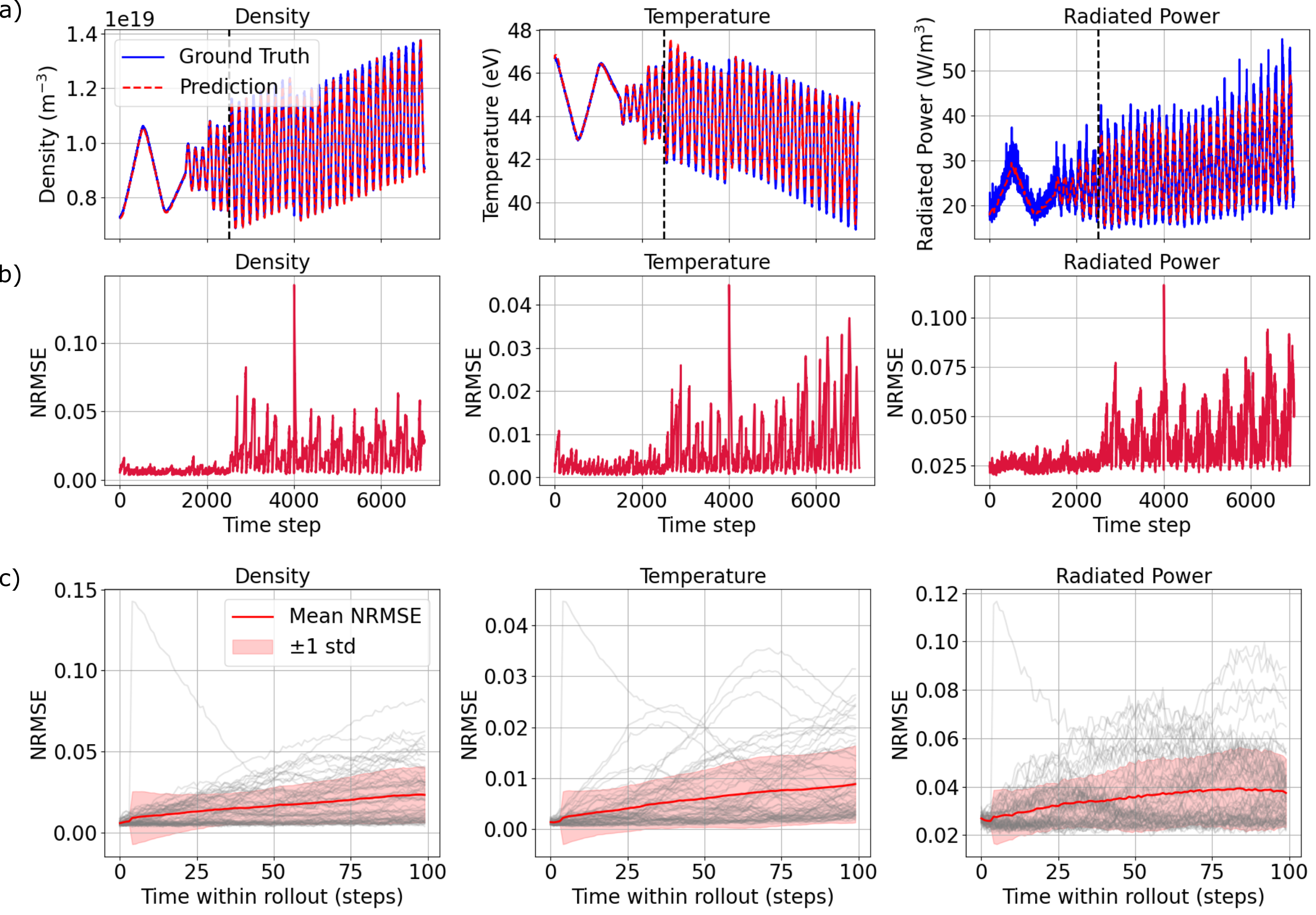}
    \caption{Temporal prediction at the outer SOL upstream near the outboard midplane (\(h=26,w=60\)) for trajectory~3 using \textit{Matey-100} with 100-step rollout. a) Ground-truth and predicted density, temperature and radiated power versus time. The vertical dashed black line marks the end of the training regime; beyond this, trajectory 3 deviates from trajectory 2 used during training. b) NRMSE versus time for the three variables. c) NRMSE versus rollout horizon: red curves denote the mean and the shaded regions indicate \(\pm1\) STD, while gray curves show the errors for individual 100-step segments. The training–testing transition is evident in panels a) and b); although errors grow with rollout horizon, they remain below 4\% on average in panel c).}
    \label{fig:1D_errors_time}
\end{figure}

To assess temporal performance, Figure~\ref{fig:1D_errors_time} presents in panel a) the ground-truth SOLPS-ITER and \textit{Matey-100}-predicted plasma states at an upstream outer SOL cell \((h=26, w=60)\), near the outboard midplane \((w=52)\) as a function of time for trajectory~3. Predictions are performed in 100-step rollout segments with the model input re-initialized to the ground truth at the start of each segment.
The vertical black dashed line marks the end of the training regime, beyond which the test trajectory~3 no longer overlaps with the training trajectory~2. 
Panel b) displays NRMSE over time, and panel c) shows NRMSE as a function of rollout step. In panel c), the red line denotes the mean NRMSE across the segments, the shaded region represents \(\pm1\) STD, and the gray lines correspond to individual 100-step segments. 
As expected, errors within the training regime (\(t<2500\) in panels a) and b)) are substantially lower than those in the test regime, \(t\geq2500\). The error oscillations arise from the periodic re-initialization with ground truth. Overall, temperature exhibits the lowest NRMSE and radiated power the highest, though still within reasonable limits as most errors are below 7.5\%. Across all variables, the largest errors occur around \(t=4000\), coinciding with an abrupt change in the density and temperature fields. Panel c) further shows that errors tend to increase gradually with longer rollouts: the mean segment-wise NRMSE at 100 steps is approximately 2\% for density, 1\% for temperature, and 4\% for radiated power, demonstrating high predictive accuracy. The mean NRMSE for radiated power is not strictly monotonic, potentially due to the interplay between model prediction error and fluctuations in the ground-truth state. 

\begin{figure}
    \centering
    \includegraphics[width=0.85\linewidth]{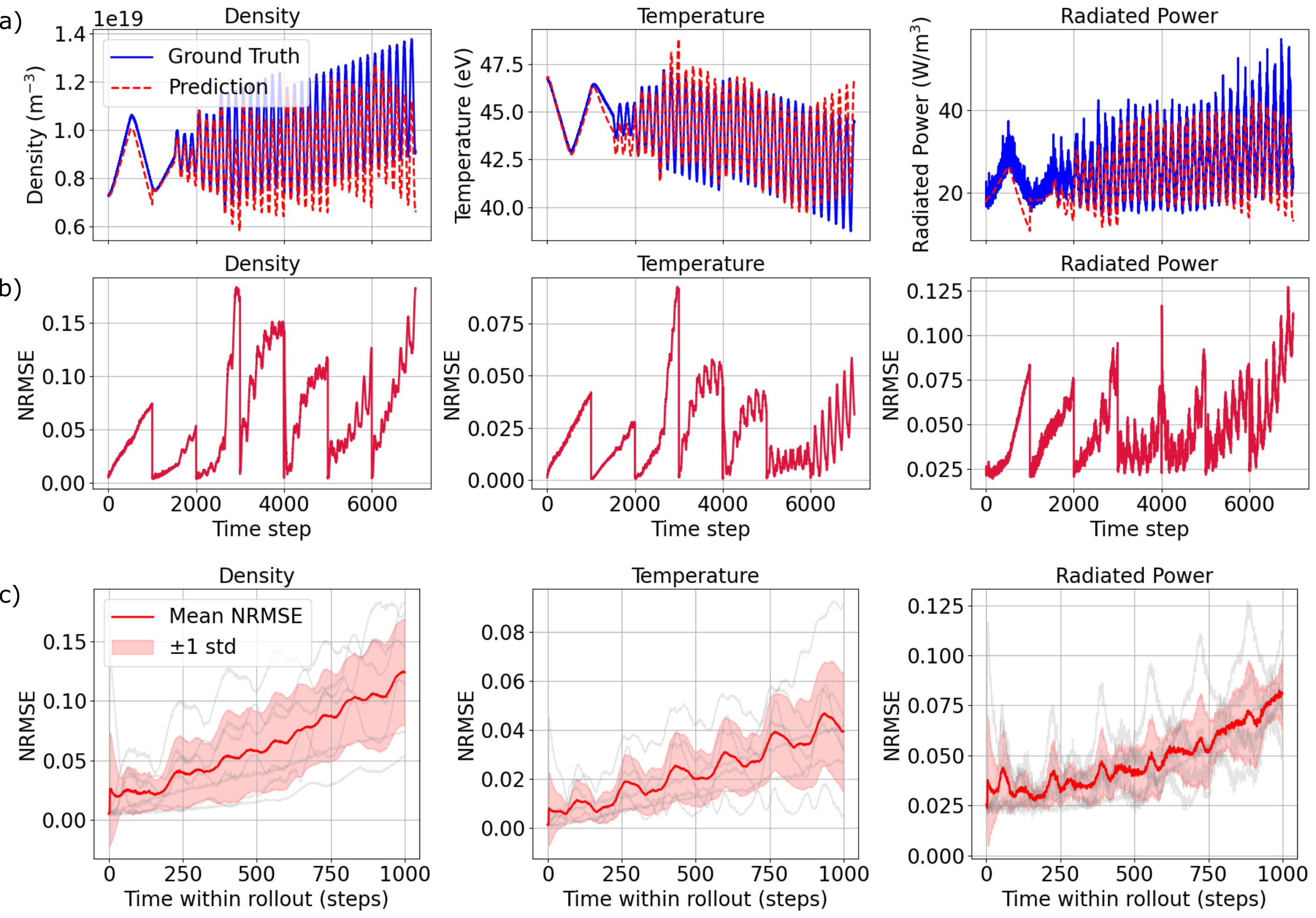}
\caption{Temporal prediction at the outer SOL upstream near the outboard midplane (\(h=26,w=60\)) for trajectory~3 using \textit{Matey-100} with 1000-step rollout. a) Ground-truth and predicted density, temperature and radiated power versus time. b) NRMSE versus time for the three variables. c) NRMSE versus rollout horizon: red curves denote the mean and the shaded regions indicate \(\pm1\) STD, while gray curves show the errors for individual 1000-step segments. Errors increase with longer rollouts; while the model captures short-horizon oscillations accurately, reproducing long-horizon trends remains challenging.
}
\label{fig:1D_time_errors_lt1000}
\end{figure}

The non-monotonicity becomes more apparent in the more challenging longer-horizon case shown in Figure~\ref{fig:1D_time_errors_lt1000}, which presents the 1000-step rollout results. Even with $10\times$ longer inference extrapolation than used during training, \textit{Matey-100} maintains strong performance, with mean segment-wise NRMSE below 13\% for density, below 5\% for temperature, and below 9\% for radiative power.
In panel c), errors within each 1000-step segment exhibit a steady increase with an oscillatory component superimposed. This pattern suggests that the model captures short-horizon dynamics while drifting in longer-term evolution. For example, the ground-truth density exhibits an increasing trend, whereas the model predicts a decreasing trend; ground-truth temperature decreases over time, but the model can capture this trend only in some segments and instead predicts either an increasing or nearly stagnant trend in others. These long-horizon discrepancies are less pronounced for radiated power, suggesting the model is comparably more robust for this variable over long-horizon prediction, even though radiated power still exhibits the highest NRMSE among the three quantities. One plausible explanation is that the ground-truth radiated power state is noticeably noisier than density and temperature. Radiated power depends on the product of electron density, impurity density (carbon in these simulations), and the impurity cooling rate, which is a nonlinear function of the electron temperature. This nonlinearity amplifies the noise in electron density and temperature when propagated into radiated power. In addition, the carbon impurity density is much lower than the main ion or electron density, causing larger Monte Carlo (MC) sampling noise. Together, nonlinear noise amplification and increased MC noise lead to substantially noisier ground-truth radiated power than for density or temperature. This noise may partially slow drift errors and regularize the surrogate's long-horizon behavior. 

Overall, \textit{Matey-100} retains predictive capability well beyond the training regime for the temporal evolution near the outboard midplane, while also highlighting the challenges of accurately capturing long-horizon dynamics in the plasma state. The relative robustness for the noisier radiated power state suggests a natural direction for future work: improving robustness in long-horizon prediction by explicitly introducing noise during training, for example through adversarial training. 

\subsubsection{1D spatial profiles at selected locations}

\begin{figure}
    \centering
    \includegraphics[width=0.95\linewidth]{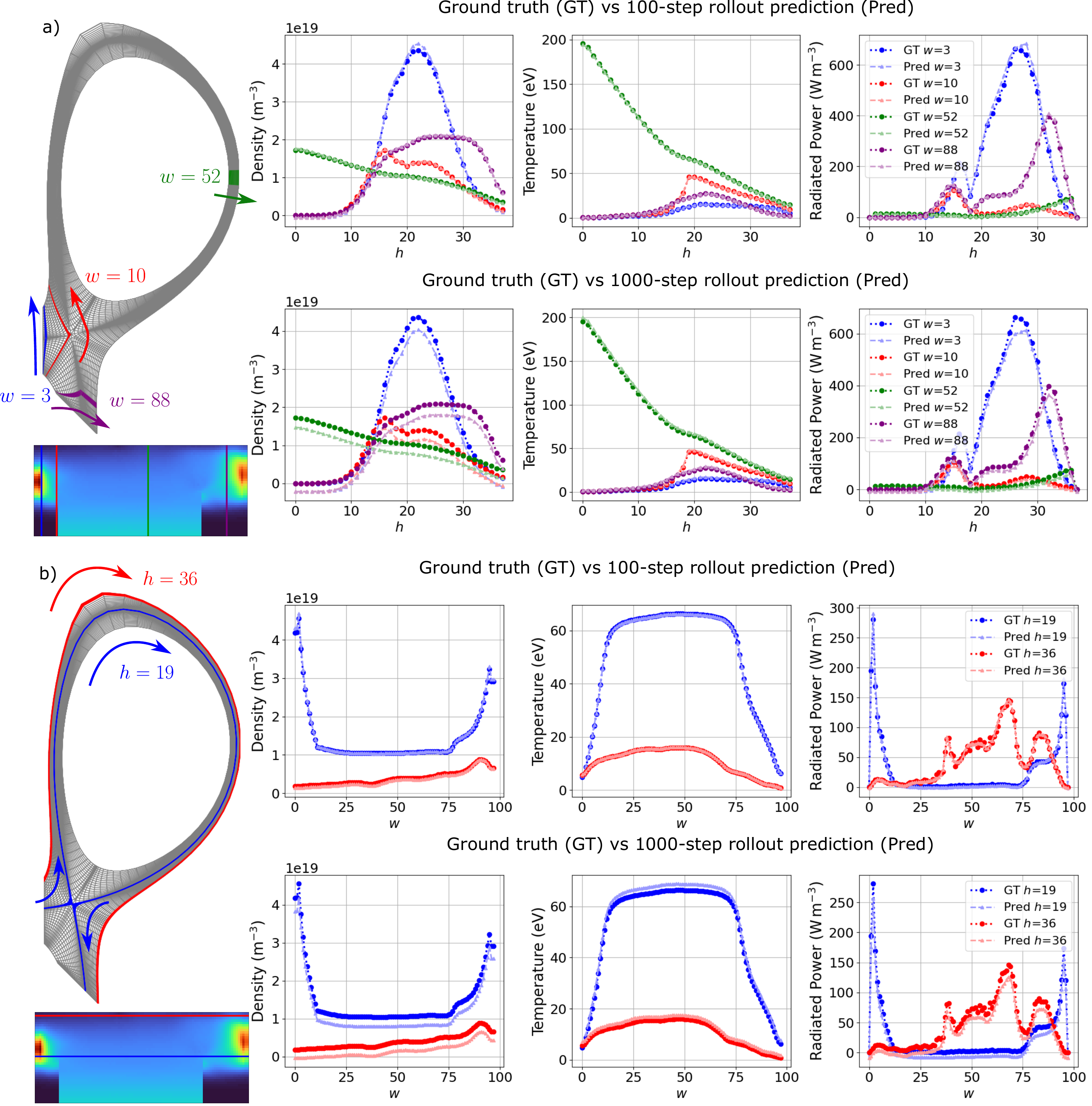}
    \caption{Density, temperature, and radiated power spatial profiles comparing SOLPS-ITER ground truth (GT) with \textit{Matey-100} predictions (Pred) for trajectory~3 at \(t=7\)~s.
    a) Plasma state variables are plotted along the radial direction ($h-$axis, vertical in the rectangular computational domain) at four locations, \(w = 3, 10, 52, 88\). The corresponding cells in the physical domain are indicated with the same colors; arrows show the slice directions. Dark lines with circles show GT and light lines with triangles represent predictions. The top and bottom rows display a 100-step rollout starting at \(t = 6.9\) s and a 1000-step rollout starting at \(t = 6.0\) s, respectively.
    b) Same comparison, but for two radial locations \(h = 19\) and \(h = 36\), along the poloidal direction ($w-$axis, horizontal in the rectangular domain). The 100-step rollout closely matches the GT, while the 1000-step rollout shows mild underprediction of density and overprediction of temperature. }
    \label{fig:1D_errors_space}
\end{figure}

Figure~\ref{fig:1D_errors_space} presents 1D profiles of predicted (Pred) and ground-truth (GT) density, temperature, and radiated power along selected spatial slices for trajectory~3 at \(t=7\)~s. The left panel marks the slice locations in both the poloidal physical domain and the rectangular computational domain.
In panel~a) (top two rows), we consider four vertical slices (along $h-$axis) at grid indices \(w=3,10,52,88\): near the left side of the divertor (blue), passing by the X-point (red), near the outboard midplane (green), and in the bottom-right region of the divertor (purple). Arrows indicate the slice directions. The top and bottom rows in panel~a) show predictions from \textit{Matey-100} with a 100-step rollout (starting at 6.9 s) and a 1000-step rollout (starting at 6.0 s), respectively. 
GT profiles are displayed in dark lines with solid circles, and predictions in lighter lines with triangles, with colors matching the slice locations. 
Panel~b) presents the same comparison for two horizontal slices (along $w-$axis) at \(h=19\) (blue; separatrix) and \(h=36\) (red; near the outer edge). 
Across panels~a) and b), the 100-step rollout closely matches the GT for all variables and slices. The 1000-step rollout remains qualitatively reasonable but exhibits reduced quantitative agreement. The model systematically underpredicts density and overpredicts temperature. Temperature is generally predicted more accurately than density: the model mostly captures the profile shapes but with a shifted magnitude in some regions. The 1000-step rollout also produces non-physical negative values, most notably for density and radiated power. Future work will address this by enforcing physical range constraints, e.g., non-negativity.

\subsubsection{Attention maps}

\begin{figure}
    \centering
    \includegraphics[width=\textwidth]{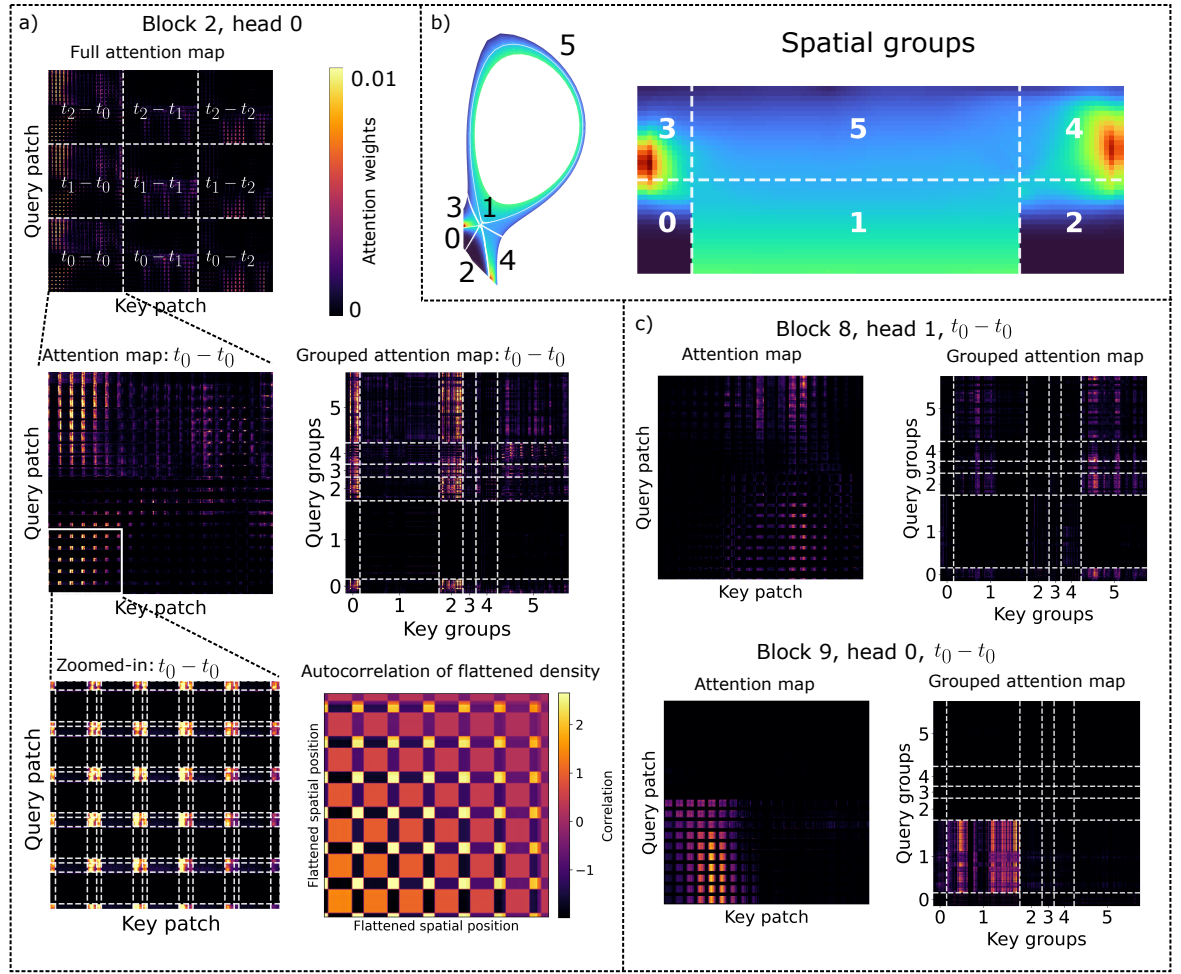}
    \caption{Representative attention maps from \textit{Matey-100} model for an input at \(t=6.9\)~s. a) block~2, head~0. Top: full $L\times L$ attention matrix, showing a distinct \(3\times3\) grid pattern corresponding to interactions among the three input snapshots at \(t_{0},t_{1}\) and \(t_{2}\). Middle: Zoomed-in view of the self-attention from \(t_{0}\) to itself (left) and its grouped attention map (right); white dashed lines indicate the group boundaries. Bottom: Further magnified region from the lower-left corner of the attention map, alongside the autocorrelation of the flattened ground-truth density field for the same spatial region. 
    b) Grouping of the computational domain into regions 0–5 in both poloidal and rectangular coordinates. Groups 0 and 2 correspond to the PFRs, group 1 to the core, and groups 3–5 to the SOL. c) Additional examples of attention maps and their grouped versions for (blocks 8, head 1) and (block 9,  head 0). All attention maps share the same colorbar from the top right of panel a). While the attention maps exhibit alignment with the defined groups, the patterns vary across examples, limiting generalizable conclusions.
}
    \label{fig:attention_figure}
\end{figure}
To better understand \textit{Matey-100} predictions, we analyze the attention maps produced by the ViT backbone. The model has 12 attention blocks, each with 3 heads, where each head outputs an \(L \times L\) attention matrix, where $L$ is the sequence length. In our setting, \(L=2793 (=T/p_t\times H/p_h\times W/p_w=3\times 19\times 49)\), yielding \(2793^{2}\) values, making direct inspection impractical. We therefore focus on selected regions to obtain more interpretable insights into the internal mechanisms of transformers.

Figure~\ref{fig:attention_figure} presents representative attention maps from \textit{Matey-100} for an input at \( t = 6.9~\mathrm{s} \) from trajectory 3. The top panel of Figure~\ref{fig:attention_figure}a) shows the full attention matrix, corresponding to attention block~2 and head~0. A distinct \(3\times3\) grid pattern is visible, reflecting the interactions among the \(T=3\) input snapshots, \( t_{0} \), \( t_{1} \), and \( t_{2} \). Moving down the left column,  the middle subplot magnifies one section of this grid, corresponding to the self-attention within \( t_{0} \), and the bottom subplot further zooms into the lower-left corner of the attention map.
Notably, the attention map exhibits pronounced patching patterns, even though the model is trained to learn all-to-all attention purely from the data. The bottom-right subplot of Figure~\ref{fig:attention_figure}a) shows the autocorrelation of the flattened ground-truth density field, and the learned attention closely mirrors this correlation structure. The high-attention patches primarily come from the PFR region. Although core-to-core regions show high autocorrelation, they have low attention in this example, but can exhibit high attention in other cases (e.g., Figure~\ref{fig:attention_figure}c), bottom). 

To further analyze the patching behavior, we partition the domain into physical meaningful groups, as shown in Figure~\ref{fig:attention_figure}b) (in both physical and computational domains). Groups~0 and~2 correspond to the PFRs, group~1 represents the core, and groups~3–5 correspond to the SOL. 
We then re-aggregate the attention map based on group ID by permuting the rows and columns. The middle-right subplot of Figures~\ref{fig:attention_figure}a) presents the resulting map, with group IDs on both axes; group boundaries are highlighted by the white dashed lines. The attention weights align clearly with these group boundaries: key groups~0 and~2 show strong attention with query groups~0,~2,~3,~4, and~5, whereas group~1 (the core) maintains consistently low attention both within itself and toward other regions due to plasma confinement across the separatrix.



Although the correlation-based observation is intriguing, it is not consistent across all blocks and all heads. Figure~\ref{fig:attention_figure}c) presents two additional examples of attention maps and their grouped counterparts. The top row corresponds to block~8, head~0, showing self-attention within \(t_0\), and the bottom row shows block~9, head~0, for the same \(t_0\) self-attention. The top example illustrates a more complex pattern, with high attention scores distributed across multiple regions and less distinctly defined boundaries; nevertheless the separation between the core (group~1) and surrounding regions remains apparent. In contrast, the bottom example exhibits a strong, but qualitatively different, alignment between attention and the spatial groupings: group~1 displays dominant self-attention and minimal interaction with other regions. Together, these results point to rich, heterogeneous inter-group correlations.

Interpreting these attention mechanisms remains challenging. 
Different heads and blocks can attend to distinct aspects of the input---some capturing broad global structures and others emphasizing localized features---therefore the physical relationships encoded by single attention map may not generalize across heads and blocks. 
In our experiments, the largest variability occurs across blocks, followed by heads, with comparatively smaller variation across the \(T=3\) input time steps. Despite this complexity, one pattern is consistent: the attention scores are mostly correlated to the spatial group partitioning. In particular, regions corresponding to the PFRs (groups 0 and 2) and the core (group 1) usually exhibit distinct behaviors, suggesting that learned attention---though operating in a high-dimensional representation space---remains shaped by the spatial structure and dynamical coupling of the physical system. Related work has attempted to encode such structure directly into DL models; for example, Zhang et al.~\cite{zhang2025calculation} imposed a prescribed attention map by group connectivity. This constraint may be too restrictive, potentially limiting the expressivity and generalizability of transformers for capturing the richer correlation patterns observed here.


\subsection{Generalization to the more challenging trajectory 3x}
\label{subsec:generalization_to_3x}

\begin{figure}
    \centering
    \includegraphics[width=\textwidth]{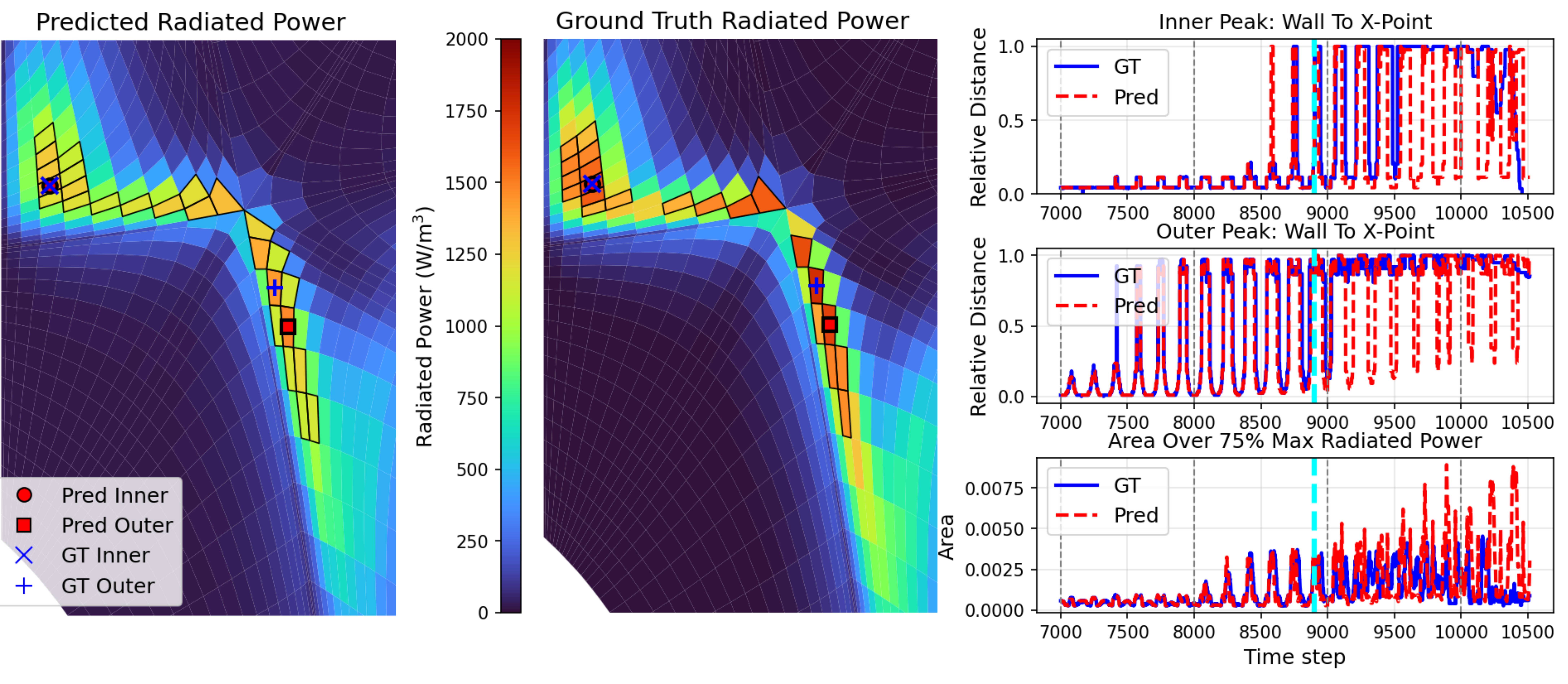}
    \caption{Radiated power for time \(t=7-10.5\)~s on trajectory~3x using a 1000-step rollout of \textit{Matey-100}. Left: predicted radiated power; middle: SOLPS-ITER ground truth (GT). Both panels are zoomed around the X-point and share the same color scale. Blue markers (‘$\times$’ and ‘$+$’) indicate the peak locations in the ground truth on the inner and outer sides, while red markers (‘$\circ$’ and ‘$\square$’) denote the predicted (Pred) peaks. Black outlines denote cells exceeding 75\% of the maximum radiated power. Right: time-resolved subplots---the relative positions of the inner and outer peak locations for Pred and GT (0 = wall and 1 = X-point; top two) and the area above 75\% of the maximum radiated power (bottom). Vertical gray lines indicate model reinitialization times, and the vertical cyan line marks timestep 8900, at which the fields in the left panels are plotted. Peak radiation locations are well captured by the model until the precursor to the X-point radiator, which represents a physical regime absent from the training data.}
    \label{fig:peak_area_rad_power}
\end{figure}

To further assess the long-horizon performance of \textit{Matey-100}, we perform inference on the more challenging extended trajectory~3x, which spans the interval \(t \in [7, 10.5]~\mathrm{s}\). 
This period represents a deeply detached plasma regime, culminating in the formation of a precursor to X-point radiator, which has recently received renewed interest as a viable operational scenario for controllable plasma exhaust regimes in future fusion reactors \cite{bernert2025xpoint} \cite{pan2023xpoint}.
Figure~\ref{fig:peak_area_rad_power} presents the inference results with a 1000-step autoregressive rollout (equivalent to 1~s of physical time). The rollout starts at \(t = 7~\mathrm{s}\) and is reinitialized at \(t = 8\), \(9\), and \(10~\mathrm{s}\). 

The left two subplots compare the predicted and ground-truth radiated power fields at \(t = 8.9~\mathrm{s}\), zoomed around the X-point and shown on the same color scale. Blue ‘$\times$’ and ‘$+$’ markers indicate the ground-truth peak locations on the inner and outer divertor sides, respectively, while red circles and squares denote the corresponding predicted peak locations. For ease of comparison, all markers are plotted on both subplots. Cells exceeding 75\% of the maximum radiated power are outlined in black for both fields.
Overall, the prediction reproduces the spatial structure of the ground truth, with a slightly reduced magnitude. The inner peak is captured accurately, while the outer peak appears marginally displaced further from the X-point. These results are obtained after 900 autoregressive inference steps; the close agreement after such a long rollout highlights the stability and robustness of the MATEY prediction.


To further assess how well the model captures the dynamics, the right-most subplots summarize the peak locations (markers in the left panels) and the high-radiation area (area enclosed by the black iso-lines in the left panels) as functions of time. We focus on the regime in which the peaks move between the divertor wall and the X-point. Accordingly, the peak locations (top two subplots) are reported using a normalized distance coordinate, where 0 indicates a peak on the divertor wall and 1 at the X-point.
The bottom subplot depicts the evolution of the area exceeding 75\% of the maximum radiated power. Vertical dashed gray lines (every 1000 steps) indicate the time steps where the model is reinitialized with ground-truth data, and the cyan line marks timestep~8900, corresponding to the 2D fields shown in the left and middle panels.
During the first two 1000-step segments (7--9~s), the model tracks both the peak locations and the evolution of the high-emission area well, after which the performance begins to degrade.
This degradation is attributed to the fact that the actuator signal (beyond approximately 9~s) exceeds the range encountered during training (i.e., at the end of trajectory~3), pushing the model into an out-of-distribution regime.
Even so, up to about 9.5~s the model continues to predict the inner peak location with reasonable accuracy, albeit with a slight phase shift. Beyond this time, the ground truth peak remains near the X-point before moving into the core region, whereas the model continues to predict oscillations of the peak position between the divertor wall and the X-point. A similar trend is observed in the total area exceeding 75\% of the maximum radiated power, where the agreement with the ground truth progressively worsens beyond 9~s.
Around \(t\approx 10.45\) s, the peak radiation occurs near the X-point, forming the precursor to an X-point radiator. Further increasing the puffing rate then pushes the radiation peak into the core region, which could lead to core contamination if not sufficiently controlled. To capture this transition, future work will need to extend the training dataset to include this physical regime.


\subsection{Runtime}
\label{subsec:runtime}

\begin{figure}
    \centering
    \includegraphics[width=0.7\linewidth]{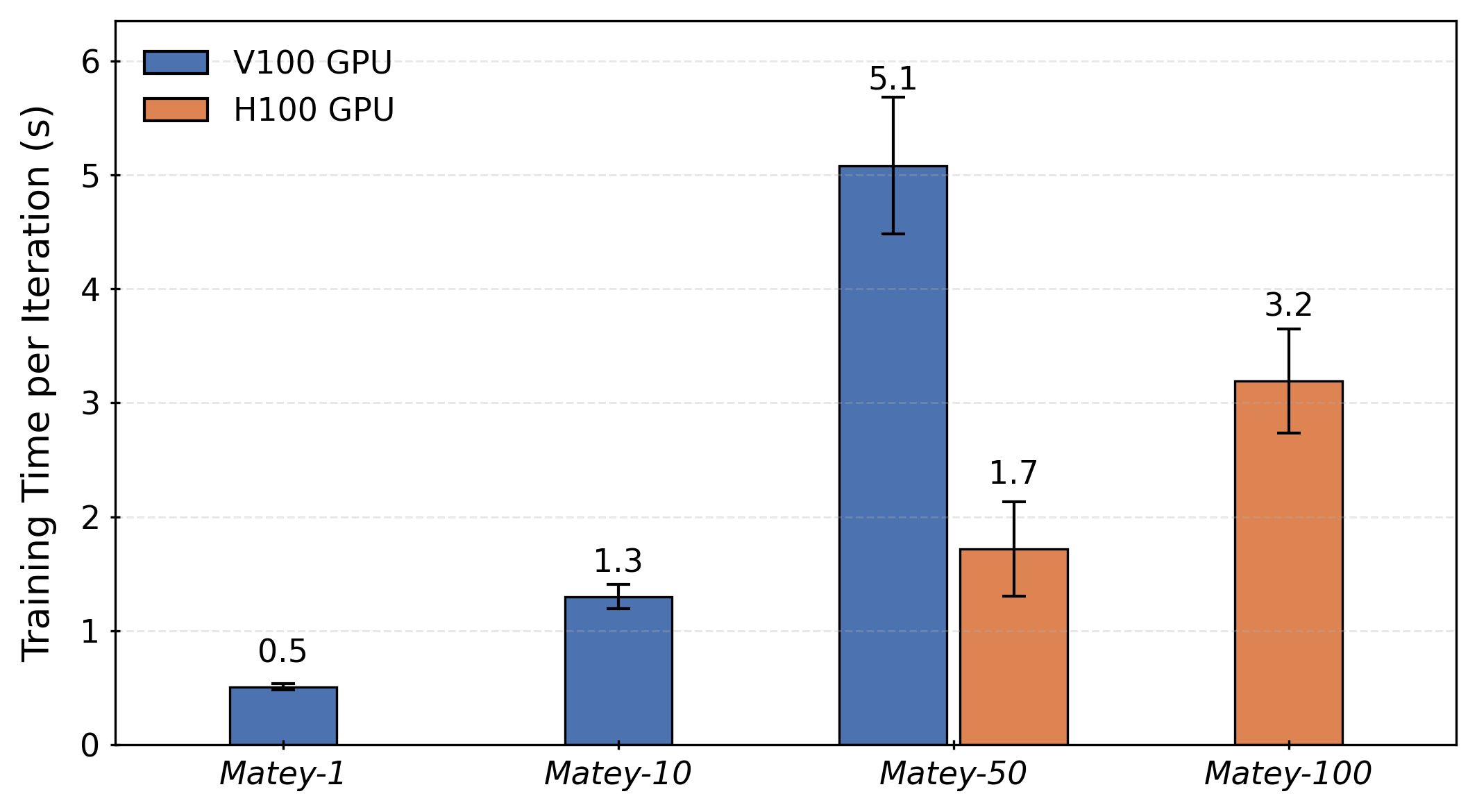}
    \caption{Average per-iteraion training time (mean\(\pm\)standard deviation) for each model, measure on two GPU types: V100 (blue) and H100 (orange). As expected, training with longer rollouts incurs higher per-iteration cost.
}
    \label{fig:train_times}
\end{figure}

Overall, our results (section \ref{subsec:global_performance}) show that training with longer autoregressive rollouts improves long-horizon predictive performance. These gains, however, come at increased computational cost. Autoregressive training requires evaluating multiple model forward passes per iteration, even though gradients are propagated only through the final pass via the pushforward trick.
Figure~\ref{fig:train_times} reports the average training time per batch iteration on V100 and H100 GPUs. For a fair comparison, we exclude the autoregressive warmup phase (the first 1000 iterations) with a restricted rollout length of \(n_{\mathrm{lead}}/2\) is ignored.
With identical model architectures and batch sizes, training \textit{Matey-50} on H100 is \(3\times\) faster than V100, while training \textit{Matey-50} takes \(10\times \) longer than training \textit{Matey-1} on V100. This scaling is expected, as each additional forward call adds overhead. 
Note that for \textit{Matey-\(n_{\mathrm{lead}}\)}, we randomly sample the actual rollout horizon uniformly between 1 and \( n_{\mathrm{lead}}\) during training, so the effective number of forward passes is smaller than \(n_{\mathrm{lead}}\). Shared costs, such as backward pass and optimizer step, further reduce the proportional slowdown at larger \( n_{\mathrm{lead}} \) values; nonetheless, \textit{Matey-100} still requires roughly twice the time of \textit{Matey-50}. 

During inference, all models use the same autoregressive procedure and therefore exhibit identical computational costs. The additional expense is thus incurred entirely offline, representing a trade-off between longer training times and improved long-horizon predictive stability. The average inference time per step (\(\Delta t=1\)~ms) (corresponding to a single forward pass) is 0.018~s on a V100 GPU and 0.009~s on an H100 GPU for a batch size of 1. For reference, the SOLPS-ITER simulation takes on the order of 30 seconds per simulated time step (\(\delta t=0.1\)~ms) using 16 MPI ranks on a Intel 2.1 GHz class CPU.

\section{Conclusion and Discussion}
\label{sec:conc-dis}
In this work, we have developed a set of transformer-based autoregressive surrogates that predict the spatiotemporal evolution of 2D divertor plasma fields from SOLPS-ITER simulations. 
\begin{itemize}
    \item  We show that increasing the rollout horizon during training improves long-horizon stability: \textit{Matey-100}, trained with 100-step rollout, reduces relative errors from 40\% (by the single-rollout model \textit{Matey-1}) to \(<10\)\%  over a full 4000-step rollout on a training trajectory, and to \(\approx 27\)\% over a 6800-step rollout on an unseen test trajectory. 
    \item We assess performance on the test trajectory (trajectory~3) from multiple aspects, including 2D field structures, time histories at the outer SOL upstream near the outboard midplane, and key 1D profiles at selected locations. Despite being trained on two trajectories, the model is able to reproduce key dynamical features and generalize to unseen inputs, underscoring the promise of attention-based architectures for modeling edge-plasma behavior. We further analyze attention maps and observe physically meaningful grouping patterns aligned with physical subdomains, suggesting the existence of correlation sparsity that could inform future more efficient attention-based model development. 
    \item To stress-test the model, we conduct inference on a more challenging trajectory~3x that enters an unseen physical regime. While the surrogate shows partial extrapolation capability and captures partially the new dynamics, it fails to catch the regime, in which the peak radiation shifts into the core region. 
    \item Lastly, we report the runtime on both training and inference. While the surrogate incurs approximation error (typically below 10\%), it achieves a time to solution that is approximately three to four orders of magnitude shorter than that of SOLPS-ITER runs, highlighting its potential for rapid design exploration.

\end{itemize}

At the same time, our results suggest several limitations that become pronounced over long horizons. Autoregressive error accumulation can lead to drift and eventually divergence from ground truth, even though longer training rollouts mitigate the issue. Model's performance degrades for regimes far from training distribution, indicating limited extrapolation capability, and extended rollouts can occasionally show non-physical behaviors. Regarding model interpretability, we probe the internal reasoning using attention scores and obtain some early insights on model behavior. However, the sheer volume and variability of attention maps across layers, heads, and time made it difficult to extract coherent physical explanations. 

As the field moves toward real-time plasma control and digital twins, accurate, robust, and uncertainty-aware surrogates will be increasingly important. 
Progress to address these issues will likely require high-quality and more diverse plasma-dynamics training data spanning broader operating regimes, boundary conditions, and control parameters, ideally with 0--5D representations across modalities and fidelities. It will also require more efficient and expressive model architectures and implementations---potentially moving toward foundation-model-style approaches ~\cite{zhang2024matey,mccabe2025walrus}---as well as tighter integration with physics-based solvers and experimental measurements. 
Finally, systematic approaches for interpretability and uncertainty quantification~\cite{kruger2024thinking}, such as Bayesian neural networks, ensemble methods, or diffusion models, remain essential for reliable prediction and control applications.

%
%

\ack{
This research is sponsored by the AI Initiative under the Laboratory Directed Research and Development (LDRD) Program of Oak Ridge National Laboratory (ORNL), managed by UT- Battelle, LLC, for the US Department of Energy (DOE) under contract DE-AC05-00OR22725. This research used resources of the Oak Ridge Leadership Computing Facility (OLCF), which is a DOE Office of Science User Facility supported under Contract DEAC05-00OR22725. The Authors acknowledge the National Artificial Intelligence Research Resource (NAIRR) Pilot and Pittsburgh Supercomputing Center (PSC) Bridges2-GPU for contributing to this research result.
This research used resources of the National Energy Research Scientific Computing Center (NERSC), a Department of Energy Office of Science User Facility using NERSC award DDR-ERCAP0030598.

This manuscript has been authored by UT-Battelle LLC under contract DE-AC05-00OR22725 with the US Department of Energy (DOE). The US government retains and the publisher, by accepting the article for publication, acknowledges that the US government retains a nonexclusive, paid-up, irrevocable, worldwide license to publish or reproduce the published form of this manuscript, or allow others to do so, for US government purposes. DOE will provide public access to these results of federally sponsored research in accordance with the DOE Public Access Plan (https://www.energy.gov/doe-public-access-plan).
}



\data{We will publish the data and code in a public repository upon acceptance of the manuscript.}
\newpage
\bibliographystyle{unsrt}
\bibliography{IOP_NucFus_template/references}

\end{document}